\documentclass[prl, twocolumn, showpacs, preprintnumbers,amsmath,amssymb]{revtex4-1}
\usepackage{graphicx}
\usepackage{dcolumn}
\usepackage{bm}
\usepackage{subfigure}
\usepackage{multirow}
\usepackage{amsmath}
\usepackage{color}  
\usepackage[font=footnotesize, justification=raggedright,singlelinecheck=false]{caption}
\begin{document}

\title{Driving dynamic colloidal assembly using eccentric self-propelled colloids}


\author{Zhan Ma}
  \thanks{These authors contributed equally.}
\author{Qun-li Lei}
  \thanks{These authors contributed equally.}
\author{Ran Ni}
 \email{r.ni@ntu.edu.sg}
 \affiliation{%
 School of Chemical and Biomedical Engineering, Nanyang Technological University, 637459, Singapore
 }

\begin{abstract}
Designing protocols to dynamically direct the self-assembly of colloidal particles has become an important direction in soft matter physics because of the promising applications in fabrication of dynamic responsive functional materials.
Here using computer simulations, we found that in the mixture of passive colloids and eccentric self-propelled active particles, when the eccentricity and self-propulsion of active particles are high enough, the eccentric active particles can push passive colloids to form a large dense dynamic cluster, and the system undergoes a novel dynamic demixing transition.
Our simulations show that the dynamic demixing occurs when the eccentric active particles move much faster than the passive particles such that the dynamic trajectories of different active particles can overlap with each other while passive particles are depleted from the dynamic trajectories of active particles.
Our results suggest that this is in analogy to the entropy driven demixing in colloid-polymer mixtures, in which polymer random coils can overlap with each other while deplete the colloids. 
More interestingly, we find that by fixing the passive colloid composition at certain value, with increasing the density, the system undergoes an intriguing re-entrant mixing, and the demixing only occurs within certain intermediate density range.
This suggests a new way of designing active matter to drive the self-assembly of passive colloids and fabricate dynamic responsive materials.
\end{abstract}

\maketitle
\section{Introduction}
Active matter are the particles capable of converting chemical and/or biological energies to drive their motion, of which the study originates from the purpose of understanding the collective self-organization phenomena in nature, like bird flocks, bacteria colonies, tissue repair, and cell cytoskeleton~\cite{revscience2012}. In recent decades, thanks to the breakthrough of particle synthesis, a number of artificial active self-propelled colloidal systems have been realized in experiments~\cite{dreyfus2005,howse2007,erbe2008,palacci2010,baraban2012,volpe2011,palacci2013,wilson2012}, 
in which the dynamics and interaction between particles can be better controlled to understand the physics of the emergent phenomena in active matter. Different from passive colloids undergoing Brownian motion due to random thermal fluctuations of the solvent, active self-propelled colloids experience an additional force due to internal energy conversion, which is random in long time with short time memories~\cite{lowen2011}.
Because of the special dynamics of active particles, they produced a number of interesting dynamic phenomena never observed in corresponding equilibrium passive matter systems, such as bacteria ratchet motors~\cite{leonado2010}, meso-scale turbulence~\cite{wensink2012}, motility induced phase separation~\cite{fily2012,redner2013,cates2013prl,speck2013,fily2014,gompper2014,cates2014theory,lowen2013,
speck2014e,bradyprl2014} and mediating emergent long range interactions~\cite{niprl2015,ray2014}. Although recently it has been shown that active colloids can serve as an medium to tune the effective interaction between passive particles~\cite{niprl2015,ray2014}, it requires the large diffusivity difference between the active and passive particles, and directing the collective assembly of passive particles using active particles remains challenging, which limits the application of using active particles in the fabrication of functional materials. 

So far, most of the studies on the collective assembly of active colloids have concentrated on the spherical or rodlike active particles, of which the self-propulsion goes through the mass center of the particle, and the dynamics is well described by a persistent random walk~\cite{howse2007}. However, due to the inhomogeneity of activity surface modification or shape anisotropicity of the particle~\cite{debnath2016diffusion,soto2014self,niu2017self,ni2017hybrid} the random self-propulsion on active particles can be eccentric, and it was found that the eccentric self-propulsion, due to the shape anisotropy of active swimmers, can significantly change the dynamics of single active particle, such as inducing the circular-like  active motion~\cite{circuleswim} and gravitaxis~\cite{hagen2014gravitaxis,wolff2013sedimentation}. Here we found that if one put these eccentric active particles into the suspension of passive colloids, the eccentric active particles can collectively push the passive colloids together to form a dense large cluster, and the system demixes.
More interestingly, we find that with certain active-passive compositions, by increasing the density of the system, the active-passive mixture undergoes an intriguing re-entrant mixing transition, and the demixing only occurs within some intermediate density range. The re-entrant behaviour offers new possibilities of using active colloids to help the fabrication of responsive functional materials.

\begin{figure*}[t!]
\centering
		\resizebox{180mm}{!}{\includegraphics[trim=0.0in 0.0in 0.0in 0.0in]{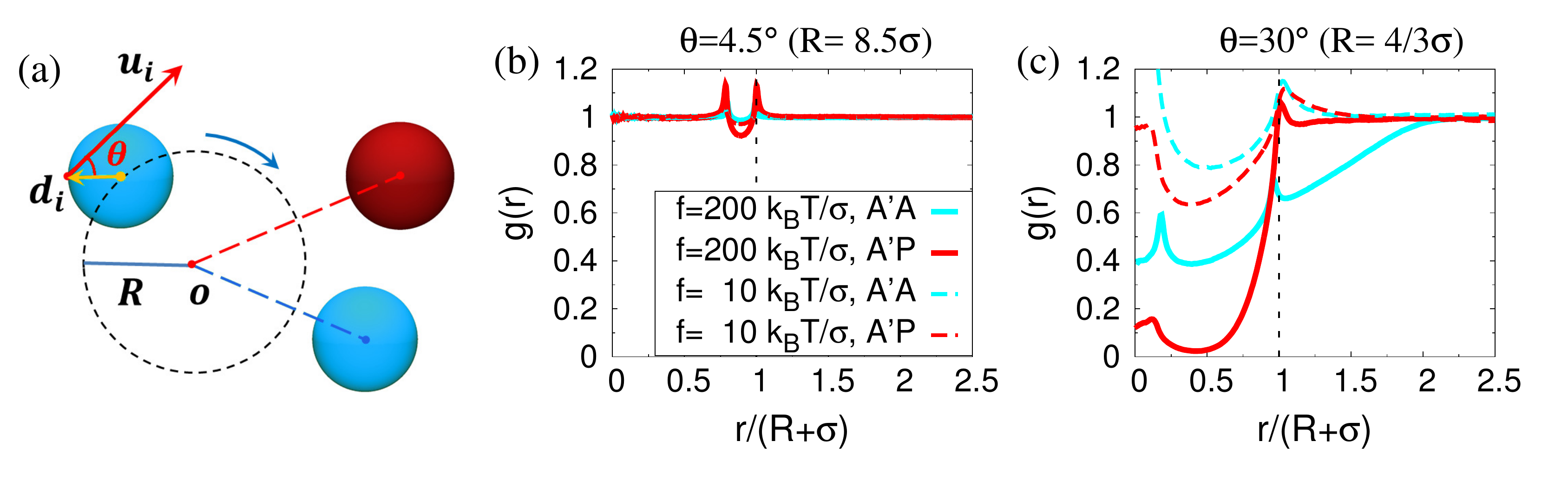} } 
\caption{\label{fig1} {\bf Illustration of the binary passive-EABP mixture.}
(a) The schematic illustration of EABPs (blue) and passive particles (red) in the system. (b,c) The radial distribution $g(r)$ of EABPs and passive particles (mass center) with respect to the instant center of circular trajectory of the reference EABP in the infinitely dilute limit for $\theta= 4.5 ^{\circ}$ (b) and $30^{\circ}$ (c) with $ f\sigma /k_BT =10$ and 200.}
\end{figure*}

\section{Model}
In experiments, eccentric active colloids are typically anisotropic~\cite{hagen2014gravitaxis,circuleswim}, while one can make the particles highly charged such that the interaction between them does not affect the orientation~\cite{liubin2014}. Therefore, focusing on the effect of eccentricity, we consider a 2D binary mixture of $N = N_p + N_a$ colloidal hard spheres with $N_p$ and $N_a$ the number of passive and active particles, respectively. 
As shown in Fig.~\ref{fig1}a, the eccentric self-propelled hard sphere is modelled as an eccentric active Brownian particle (EABP) with a spherical diameter $\sigma$, and a constant self-propulsion acting on the colloidal surface, which has an eccentric angle $\theta$ from the center of the particle. 
This eccentric self-propulsion induces a constant torque $\mathbf{\Omega}_i$ on particle $i$ given by
\begin{eqnarray}
{\bm \Omega}_i = f_i  \cdot \textbf{d}_i \times \textbf{u}_i,
\end{eqnarray}
where $\mathbf{d}_i$ is the vector connecting the mass center of the particle and the location of the self-propulsion with $f_i$ and $\mathbf{u}_i$ the strength and orientation of the self-propulsion on particle $i$, respectively. Here we use $|\mathbf{d}_i| = \sigma/2$. For passive particles, $f_i = 0$. The strength of the active torque is $|\mathbf{\Omega}_i| = f_i |\mathbf{d}_i| \sin \theta$, and we define $\theta$ as the eccentricity of the self-propulsion.
In our 2D system, $\mathbf{\Omega}_i$ is a constant vector always pointing towards the inside or outside of the plane, and for the sake of simplicity we choose $\theta > 0$ and $\mathbf{\Omega}_i$ points to the inside of the plane. Even though the particles are driven and energy is continuously supplied to the system, we assume the solvent to be at an equilibrium temperature $T$.
The motion of particle $i$ with position $\mathbf{r}_i$ and orientation $\mathbf{u}_i$ is described via the overdamped Langevin equation given by
\begin{eqnarray}
\dot{\textbf{r}}_i(t) &=& \frac{D_0}{k_BT} \left[-{\nabla}_i U(t) + f_i\cdot \textbf{u}_i(t)\right] + \sqrt{2D_0}{\bm \xi}_t(t),\\
\dot{\textbf{u}}_i(t) &=&   \left[ \frac{D_r {\bm \Omega}_i}{k_BT} + \sqrt{2D_r}{\bm \xi}_r(t)\right] \times {\textbf{u}}_i(t),
\end{eqnarray}
where $D_r$ is the rotational diffusion constant of the particle with $k_B$ the Boltzmann constant.  According to the Stoke-Einstein relationship~\cite{ni2013natcom}, we assume $D_r=3 D_0/\sigma^2$ with $D_0$ the short-time translational self-diffusion coefficient of the sphere.
${\bm \xi}_t(t)$ and ${\bm \xi}_r(t)$ are Gaussian thermal noises with zero mean and unit variance. 
For a given $\theta$, without the thermal noises, the EABP performs circular motion with a radius $R=|\dot{\textbf{r}}|/|\dot{\textbf{u}}| = 2\sigma/(3\sin \theta)$. Here $R$ reaches the minimal value of $2\sigma/3$ at $\theta = \pi/2$, and it is due to the fact that the torque is induced by the eccentricity of the self-propulsion, which is true for shape anisotropic active swimmers in experiments.
$U$ is the total potential energy of the system, which is the sum of all pairwise interactions $U(r_{ij})$ given by the Weeks-Chandler-Andersen (WCA) potential to mimic the colloidal hard-sphere interaction~\cite{wca}
\begin{eqnarray}
\frac{U(r_{ij})}{k_BT}=\left\{
\begin{array}{lr}
4\left[\left(\frac{\sigma}{r_{ij}}\right)^{12}-\left(\frac{\sigma}{r_{ij}}\right)^{6}+\frac{1}{4}\right] & (r_{ij}<2^{1/6}\sigma)\\
0& (r_{ij}>2^{1/6}\sigma)
\end{array}
\right.
\end{eqnarray}
where $r_{ij}$ is the center to center distance between particle $i$ and $j$.

\begin{figure*}
\centering
		\resizebox{180mm}{!}{\includegraphics[trim=0.0in 0.0in 0.0in 0.0in]{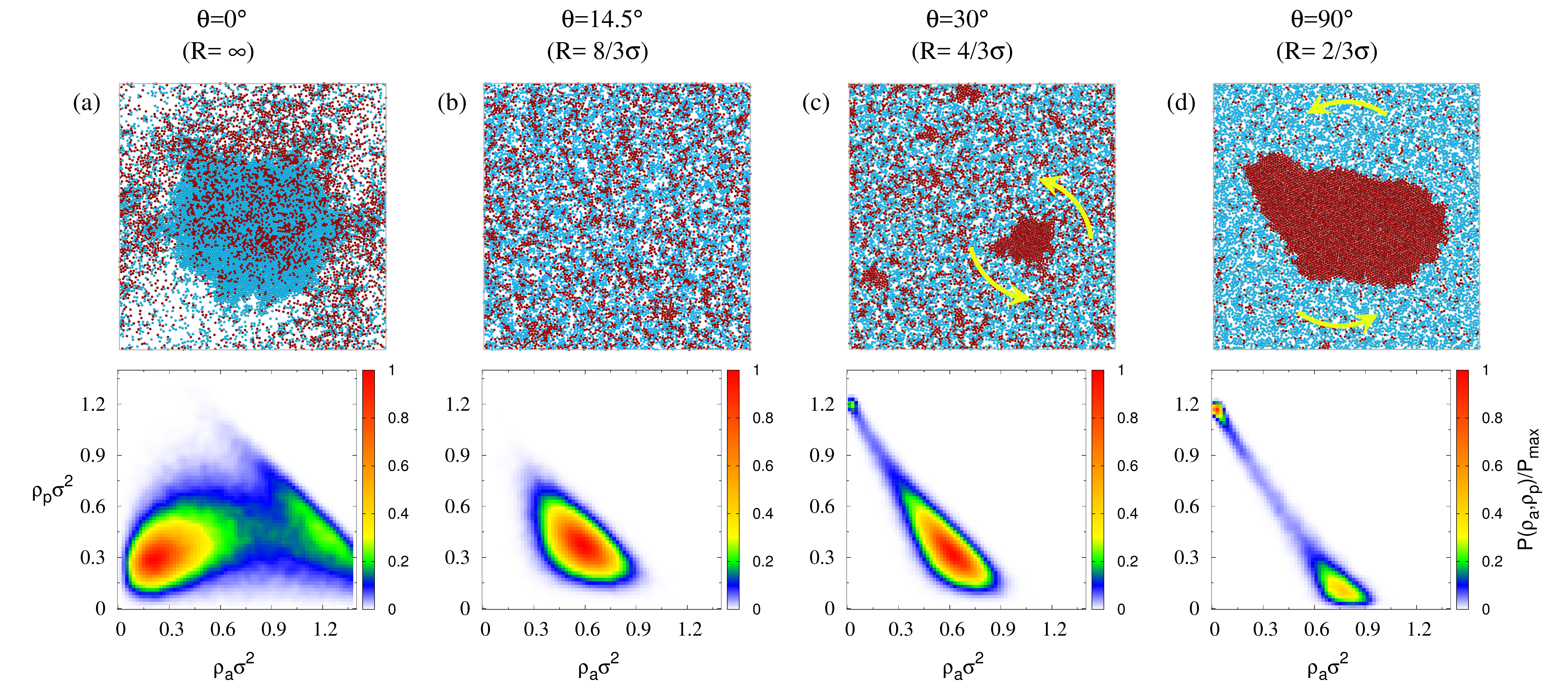} } 
\caption{\label{fig2} {\bf Phase behavior of the 2D passive-EABP mixtures with various eccentricity.} Typical snapshots (upper panels) and the reduced probability distribution of local density $P(\rho_a,\rho_p)/P_{\max}$  (lower panels) in the binary mixture of passive particles (red) and EABPs (blue) at $\rho \sigma ^ 2 = 0.891$, $x_p=0.4$, and $f\sigma/k_B T =200$ with various eccentricity: (a) $\theta=0\ ({\rm R}= +\infty)$; (b) $\theta=14.5^{\circ}\  ({\rm R}= 8/3\sigma)$; (c) $\theta=30^{\circ}\  ({\rm R}= 4/3\sigma)$; (d) $\theta=90^{\circ}\  ({\rm R}= 2/3\sigma)$. Here $\rho_a$ and $\rho_p$ are the local density of EABPs and passive particles, respectively, with $P_{\max}$ the maximal probability in $P(\rho_a,\rho_p)$. Movies of the systems can be found in Supplementary S1. The arrows indicate the direction of rotation of the passive particle cluster.}
\end{figure*}

\section{Results}
We first study the infinitely dilute mixture of passive and EABPs with different $f$ and $\theta$, in which we simulate two particles and the system size is large enough to exclude any effect of the periodic boundary condition. 
In Fig.~\ref{fig1}b,c, we plot the radial distribution functions $g_{A'A(P)}(r)$ of the system, in which $r$ is the distance between the instant center of the circular trajectory of an reference EABP and the mass center of another EABP (passive) particle.
One can see that when the eccentricity is small, i.e. $\theta= 4.5^{\circ}$, $g_{A'A}(r)$ and $g_{A'P}(r)$ are very similar to each other and do not change much with increasing $f$, and both EABPs and passive particles can enter the circular trajectories of other EABPs. With increasing the eccentricity to $\theta = 30^{\circ}$, $g_{A'A}(r)$ becomes larger than $g_{A'P}(r)$ at distance $r < R+\sigma$, and the difference becomes more substantial with increasing $f$.
This implies that compared to EABPs, passive particles are more depleted from the zone within the circular trajectories of other EABPs, which effectively induces an ``extra short range repulsion'' between EABPs and passive particles.

To further investigate the effect of eccentricity on the collective assembly of the system, we simulate a binary mixture of EABPs and passive particles with the overall density $\rho \sigma^2 = 0.891$ and composition $x_p = N_p/(N_a + N_p) = 0.4$, where the self-propulsion of EABPs is fixed at $f\sigma/k_BT =200$, and eccentricity varies from $\theta = 0$ to $90^{\circ}$. As shown in Fig.~\ref{fig2}a, when $\theta = 0$, the system recovers the conventional binary mixture of passive and active Brownian particles~\cite{stenhammar2015a}, and phase separates into a high density liquid-like phase and a low density gas-like phase, whose densities can be obtained from the location of two peaks in the reduced probability distribution of local density in the lower panel of Fig.~\ref{fig2}a similar to Ref.~\cite{wysocki2016traveling,stenhammar2015a}. 
When increasing the eccentricity to $\theta = 14.5^{\circ}$, the gas-liquid like phase separation disappears, and the bimodal probability distribution of local density merges into a single peak in Fig.~\ref{fig2}b suggesting a homogeneous phase. Interestingly, as shown in Fig.~\ref{fig2}c, with further increasing the eccentricity to $\theta = 30^{\circ}$, a different bimodal probability distribution of the local density appears, and the system forms a dense cluster of pure passive particles with density $\rho_p \sigma ^2 \simeq 1.2$  surrounded by a binary fluid mixture with $(\rho_a \sigma^2, \rho_p \sigma^2) \simeq (0.6,0.3)$. This implies the demixing of passive particles from the passive-EABP mixture. Moreover, as the system is in 2D, the cluster of passive particles formed in the system spontaneously rotates towards the opposite rotating direction of EABPs. By further increasing the eccentricity to $\theta = 90^{\circ}$, one can observe a more pronounced passive-EABP demixing in the system, which almost completely phase separates into a high density phase of passive particles and a low density phase of EABPs.

\begin{figure}
\centering
		\resizebox{90mm}{!}{\includegraphics[trim=0.0in 0.0in 0.0in 0.0in]{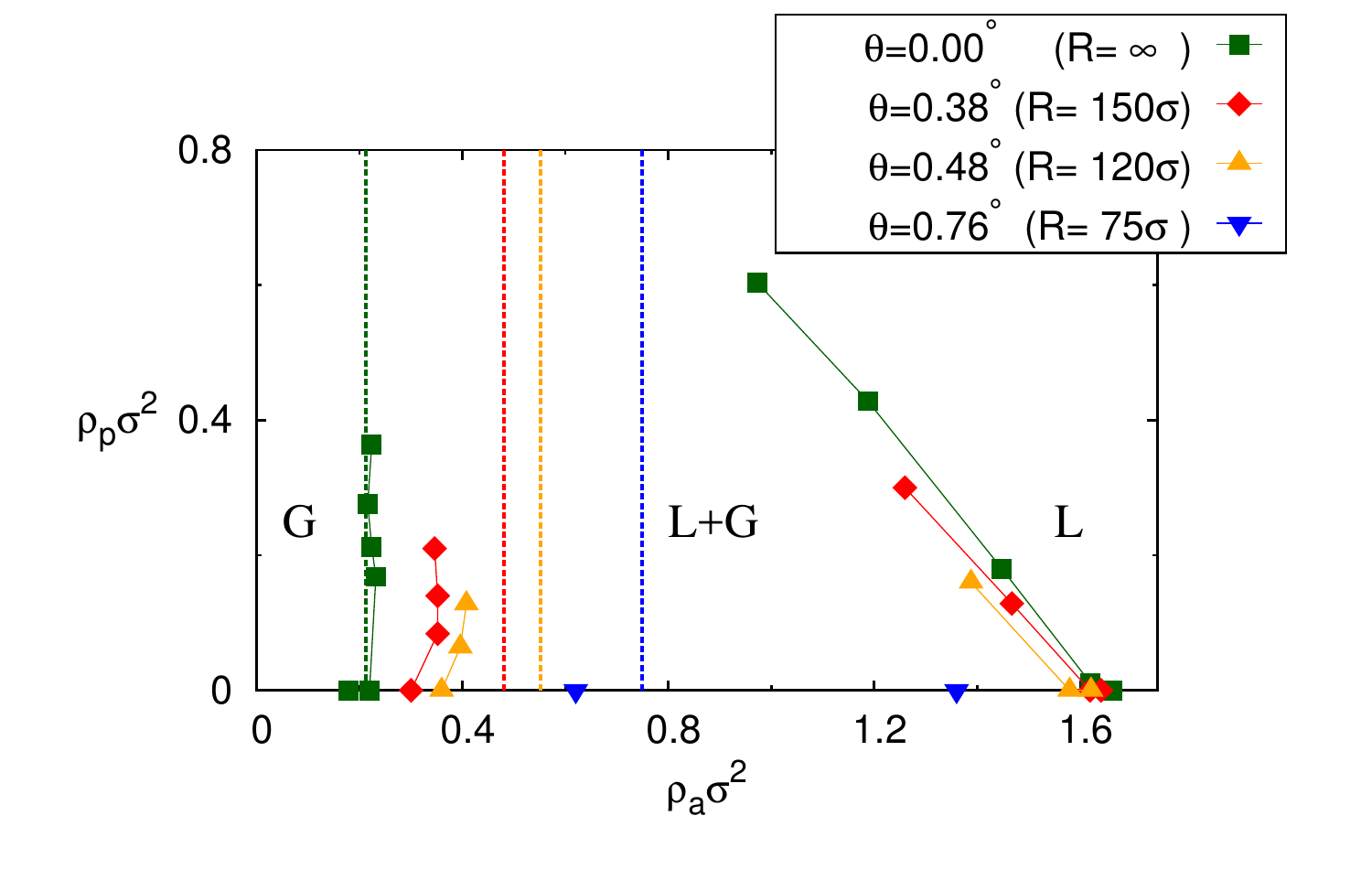} }
\caption{\label{fig3}{\bf Motility induced phase separation in passive-EABPs mixtures.} Phase diagrams of the 2D binary mixture of passive and EABPs with $f\sigma/k_B T = 200$ and weak eccentricity $\theta$ in the $\rho_a-\rho_p$ representation. The dashed lines are the theoretical prediction in Eq.~\ref{theory}.}
\end{figure}

By locating the peaks in the probability distribution of local density, like in Fig.~\ref{fig2}a,b, we construct the phase diagram for the motility induced gas-liquid like phase separation in mixture of passive and EABPs with weak eccentricity (symbols in Fig~\ref{fig3}). One can see that when $\theta$ is zero, the phase diagram features a gas-liquid like phase separation of two phases with similar passive particle densities and different active particle densities. By adding passive particles to the mixtures, the density difference between the co-existing phases becomes smaller, and the co-existing phase envelope eventually closes at the high enough density of passive particles, which agrees with Ref.~\cite{stenhammar2015a}. With slightly increasing the eccentricity $\theta$, the binodal of non-equilibrium gas phase increases dramatically, and the gas-liquid like phase envelope becomes smaller, which almost completely disappears at $\theta > 0.76^{\circ}$. To understand the disappearance of the motility induced gas-liquid like phase separation in the system, we build up a phenomenological theory following the kinetic model in Ref.~\cite{redner2013,stenhammar2015a}. In 2D, we assume that the gas-liquid interface consists of close packed EABPs~\cite{stenhammar2015a}, and the orientation of an EABP $\psi$ satisfies the Langevin equation:
\begin{eqnarray}\label{eqbd}
\frac{d\psi}{d t}=\frac{D_r \Omega}{k_B T} +\sqrt{2D_r}\xi_r,
\end{eqnarray}
by solving which we obtain the orientation correlation function $\langle \cos \Delta \psi(t) \rangle =\cos(2\pi \omega t)e^{-D_rt}$ with $\Delta \psi(t)= \psi(t)-\psi(0)$ and $\omega=D_r\Omega/2 \pi k_BT= f D_r  \sigma \sin\theta /4\pi k_BT$. When $\theta \rightarrow 0$, we have $D_r \gg \omega$ and $\langle \cos\Delta \psi(t) \rangle \simeq e^{-D_rt}$, in which the characteristic time for an active particle on the gas-liquid interface escaping to the gas phase is $\tau_{out} = 1/D_r$. On the other hand, when the eccentric torque on the active particle is very large, i.e. $D_r \ll \omega$  , the characteristic time for an active particle to escape from the surface can be approximated as $\tau_{out}=1/4\omega$, which is the time that the orientation correlation function first becomes zero. Therefore, generally, as the first approximation, we assume $\tau_{out} = 1/(D_r + 4 \omega)$, and the flux of EABPs per unit length on the gas-liquid interface escaping from the liquid phase $k_{out} = \kappa /( \sigma \tau_{out})$ with $\kappa = 4.5$ obtained from the fitting in Ref.~\cite{redner2013}. Similarly, we also obtain the flux of active particles per unit length on the phase boundary entering the liquid phase as $k_{in} = \rho_{g}^{a} v_p/\pi$, where $\rho_{g}^{a}$ is the density of active particles in the non-equilibrium gas phase with velocity $v_p = f D_0/k_BT$. By equating $k_{out} = k_{in}$, we  obtain the binodal of the gas phase 
\begin{eqnarray}\label{theory}
\rho_g^a= \frac{\pi \kappa (D_r +4\omega)}{ v_p \sigma},
\end{eqnarray}
of which the dependence on $\theta$ is shown as the vertical dashed lines in Fig.~\ref{fig3}. One can see that by slightly increasing $\theta$, the binodal of gas phase increases significantly, which qualitatively agrees with the phase boundary obtained from simulations. Essentially, this implies that the disappearance of the gas-liquid like motility induced phase separation in the passive-EABP mixture is due to the decreased orientation relaxation time by the active torque $\Omega$, which is similar to the effect of increasing $D_r$ on the motility induced phase separation~\cite{hot_cold}.

\begin{figure*}
\centering
		\resizebox{180mm}{!}{\includegraphics[trim=0.0in 0.0in 0.0in 0.0in]{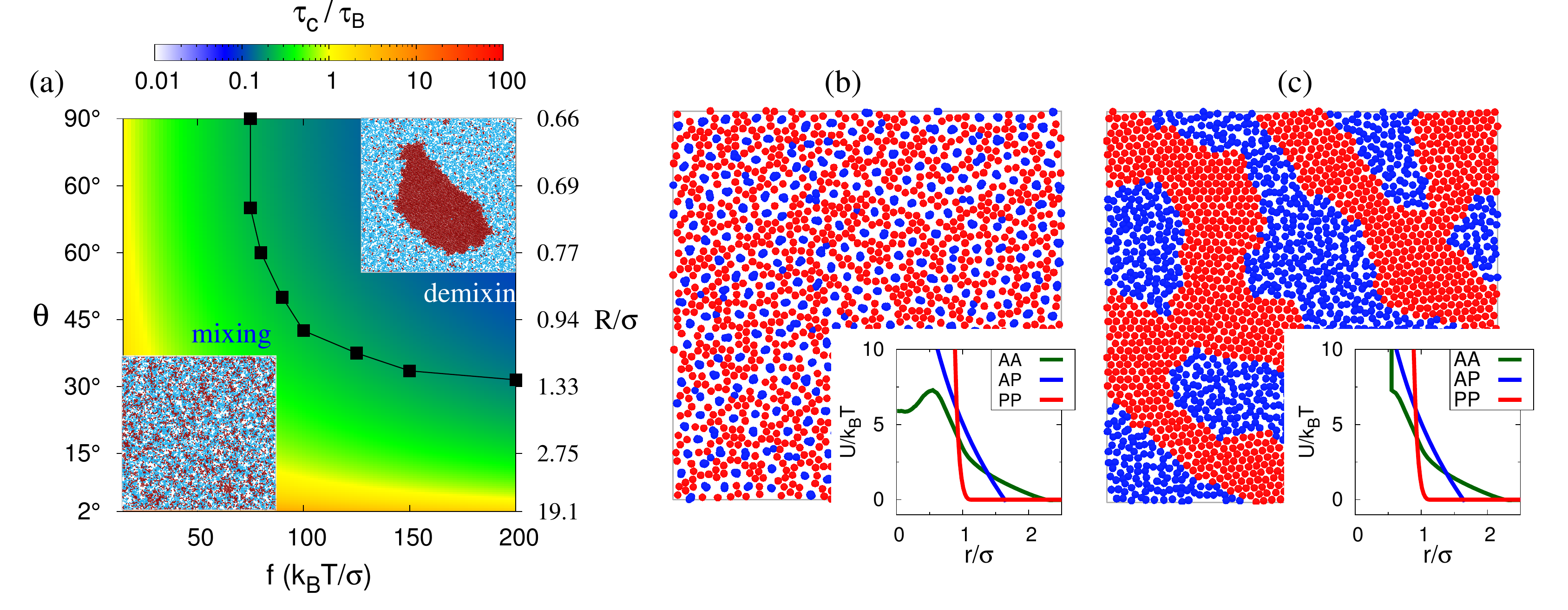} }
\caption{\label{fig4} {\bf Dynamic demixing in passive-EABP mixtures.} (a) Phase diagram of the 2D binary passive-EABP mixtures in the $f-\theta$ representation with $\rho \sigma ^ 2=0.891$ and $x_p=0.4$, where the black squares are the demixing phase boundary. The insets are typical snapshots of the mixed and demixed systems, where red and blue spheres are passive and eccentric active particles, respectively. The background represents the value of $\tau_c/\tau_B$ according to the color bar on the right. (b,c) Typical snapshots of the binary system of A (blue) and P (red) particles at $\rho \sigma ^ 2 = 0.891$ and $x_p = 0.4$ (same as in Fig.~\ref{fig2}d), and the pair interactions used are shown in the insets. Pair interactions in (b) are obtained from simulations of infinitely dilute binary passive-EABP mixtures with $f\sigma/k_BT = 200$ and $\theta = 90^{\circ}$, based on which those in (c) are modified by including an excluded volume effect between the A-A interaction.}
\end{figure*}

Additionally, as shown in Fig.~\ref{fig1}c and d, with further increasing the eccentricity to $30^{\circ}$ and $90^{\circ}$, the system undergoes a new dynamic demixing by forming a large dense cluster of passive particles surrounded by non-equilibrium binary fluids. By using the quaternion-based Rotational Brownian Dynamics algorithm~\cite{ilie2015e}, we confirm that the demixing also occurs in the 3D binary passive-EABP mixtures in Supplementary S2, where the orientation of the active torque on each EABP performs free Brownian rotation. To understand the essential physics driving this intriguing demixing in the passive-EABP mixture, we simulate the system with different combinations of $f$ and $\theta$ at $\rho \sigma^2 = 0.891$ and $x_p = 0.4$. To identify the critical $f$ and $\theta$ for the demixing, we calculate the demixing order parameter $\langle \alpha \rangle = \frac{1}{A_a} \langle (1-x_p)[ \sum_{i=1}^{N_a}{(n_{i}^a - n_{i}^p)/6N_a + 2x_p - 1}]\rangle + \frac{1}{A_p}\langle x_p [ \sum_{i=1}^{N_p}{(n_{i}^p - n_{i}^a)/6N_p - 2x_p + 1}] \rangle$, where $n_{i}^{a/p}$ is the number of EABP/passive particle neighbours in the 6 nearest neighbours of particle $i$. $A_a$ and $A_p$ are the normalization factors, which are chosen such that $\langle \alpha \rangle = 1$ when EABPs and passive particles are fully demixed, and $\langle \alpha \rangle = 0$ when the two types of particles are miscible. As a single large cluster forms when $\langle \alpha \rangle \simeq 0.2$, we use $\langle \alpha \rangle > 0.2$ as the criteria of demixing, and the resulting phase boundary is plotted in Fig.~\ref{fig4}a. We can see that with increasing the strength of self-propulsion, the critical eccentricity for driving the demixing decreases significantly. When $f\sigma/k_BT \simeq 80$, the critical eccentricity for demixing is around $\theta \simeq 90^{\circ}$, while $\theta$ drops to about $30^{\circ}$ for $f\sigma/k_BT \simeq 200$.
In the binary mixture, there are essentially two different dynamic time scales for the motion of passive particles and EABPs, respectively: (i) $\tau_B = \sigma^2/4D_0$ the average time for a passive particle diffusing for the distance of $\sigma$; (ii) $\tau_c=4\pi\sigma k_BT/(3 f D_0 \sin\theta )$ the period of circular motion of an EABP. We superpose the ratio between these two dynamic time scales $\tau_c/\tau_B$ on the phase boundary in Fig.~\ref{fig4}a. We can see that the system demixes at small $\tau_c/\tau_B$, in which the EABPs perform circular-like motion much faster than the Brownian diffusion of passive particles. As seen from our simulation of dilute systems, when EABPs move much faster than passive particles, passive particles are depleted from the dynamic trajectories of EABPs, which effectively creates a ``repulsion'' between the dynamic trajectories and passive particles. At the meanwhile, as EABPs are all moving on the same time scale, their dynamic trajectories can overlap with each other such that the ``repulsion'' between the dynamic trajectories of active particles is weaker. This is similar to the system of colloid-polymer mixtures, in which polymer random coils can to some extent overlap with each other but not with the colloidal particles inducing an stronger repulsion between colloid and polymer than that of polymer-polymer interaction. When the density is high enough, the colloid-polymer mixture demixes due to the entropy maximization of the system~\cite{colloid_polymer}, and in our non-equilibrium system of passive-EABP mixtures, similar demixing also occurs. 

To test the picture of entropy driven demixing, we assume the circular-like dynamic trajectory of an EABP as a quasi-particle, namely A particle, whose center is the instant center of the circular trajectory, and passive particles can be seen as another type of particles, namely P particles. From the simulation of infinitely dilute passive-EABP mixtures, we can extract the effective pair potentials among A and P particles, which are used as input for equilibrium simulation of the system at high densities. From the obtained pair potentials in the inset of Fig~\ref{fig4}b, one can see that the repulsion between A-A particles is softer than that between A-P particles, which is similar to the colloid-polymer mixtures~\cite{star_colloid}. However, as shown in Fig.~\ref{fig4}b, the equilibrium simulation at high density, i.e. $\rho \sigma^2 = 0.891$ and $x_p = 0.4$, does not show a demixing seen in the corresponding passive-EABP mixture in Fig~\ref{fig2}d. From the snapshot of the equilibrium simulation, we can see that many A particles overlap with each other due to the potential well between A-A particles at the short distance $r \lesssim 0.5\sigma$, and most of the blue spots are occupied by at least 4 A particles in Fig.~\ref{fig4}b. This is the artefact of the effective pair potential, as in our simulation of passive-EABP mixtures, dynamic trajectories of two EABPs can overlap and share the circular center with $R=2/3\sigma$, while because of the hard-core interaction between EABPs, it is not possible to have more than 3 active particles sharing the same circular trajectory, and this number can increase at the larger $R$. Therefore, this suggests that there should be a multi-body excluded volume term in the effective interaction among A-A particles.
If we just add an excluded volume effect at short distance, i.e. $r \lesssim 0.5\sigma$, to the A-A interaction (Fig.~\ref{fig4}c inset), the demixing between A and P particles occurs  as shown in Fig.~\ref{fig4}c. This suggests that the multi-body excluded volume effect induced by the non-equilibrium dynamics is important for the dynamic demixing of the system at high density, and by including proper excluded volume effects between dynamic trajectories of EABPs, i.e. A particles, one can indeed use effective interaction obtained from low density passive-EABP systems to qualitatively reproduce the dynamic demixing at high density. It is worth noticing that as shown in Fig.~\ref{fig4}a, when $f \rightarrow \infty$, the phase boundary is approaching certain threshold $\theta$, e.g. about $30^{\circ}$ in Fig.~\ref{fig4}a, depending on the density and stoichiometry of the system, and this can not be explained by using the ratio between two dynamic time scales $\tau_c/\tau_B$, which approaches zero at $f \rightarrow \infty$. Therefore, the physics of this threshold of $\theta$ at $f \rightarrow \infty$ remains interesting for further investigation.

\begin{figure}
\centering
		\resizebox{90mm}{!}{\includegraphics[trim=0.0in 0.0in 0.0in 0.0in]{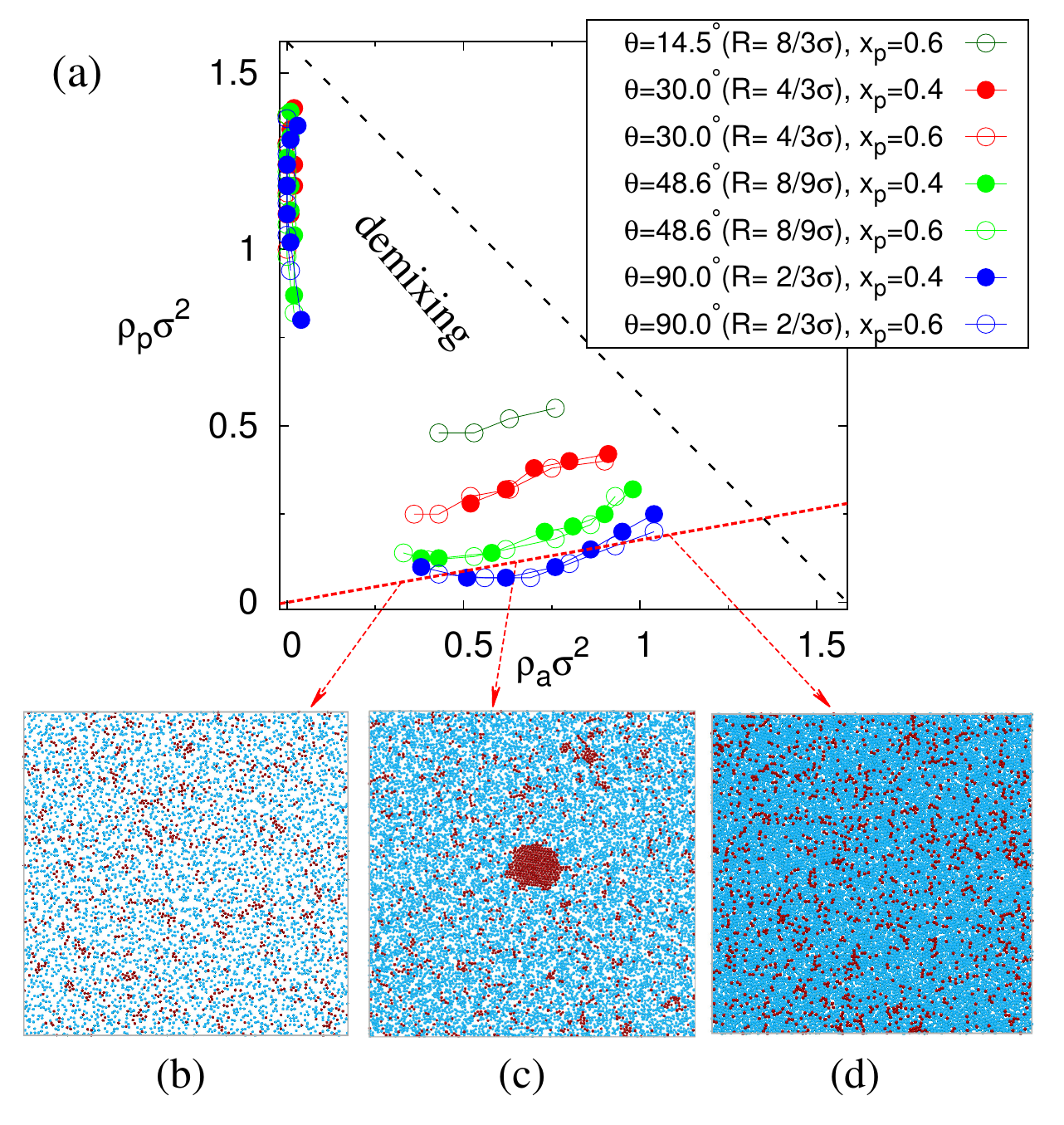} } 
\caption{\label{fig5} {\bf Dynamic demixing phase diagram.} Phase diagram of the 2D binary passive-EABP mixtures with $f\sigma/k_B T = 200$ and various eccentricity $\theta$ in the $\rho_a-\rho_p$ representation, where $x_p$ is fraction of passive particles in the simulated systems, and the dashed line indicates the close packing obtained.
(b-d): Typical snapshots of the 2D binary mixture of passive particles and EABPs with $f\sigma/k_B T = 200$ and $x_p = 0.15$ at different densities: $\rho \sigma^2=0.38$ (b), $0.76$ (c), and $1.27$ (d).
 Here the blue and red spheres are EABPs and passive particles, respectively.}
\end{figure}

Furthermore, we construct the demixing phase diagram for the binary mixture of passive particles and EABPs of $f \sigma /k_BT = 200$ with various eccentricity $\theta$ as shown in Fig.~\ref{fig5}a, in which we simulate the system of $N=N_a + N_p = 5,000 \sim 10,000$ particles to ensure the system size large enough to eliminate the finite size effect. When $\theta \gtrsim 14.5^{\circ}$, the dynamic demixing occurs in the system. One can see that with increasing the eccentricity $\theta$, the EABP rich phase boundary significantly moves to the lower fraction of passive particles, while the composition in the passive particle rich phase once formed is almost independent with eccentricity.
We also varies the passive particle composition $x_p$ in the system to ensure that similar to equilibrium demixing, the obtained phase boundary is independent with $x_p$ and the lever rule still holds. More interestingly, as shown in Fig.~\ref{fig5}a, the density of passive particles in the EABP rich phase changes non-monotonically with increasing the density of EABPs $\rho_a$, and develops a minima. Therefore, as shown in Fig.~\ref{fig5}b-d, by fixing the passive particle composition at $x_p = 0.15$ with $\theta = 90^{\circ}$, and increasing the density of the binary mixture from $\rho \sigma^2 = 0.38$ to 1.27, the system undergoes an interesting re-entrant mixing, which demixes at the intermediate density, i.e. $\rho \sigma^2 = 0.76$. 
This re-entrant mixing at high density is due to the fact that the system is too crowding such that EABPs can not perform proper circular-like motion, and the dynamic entropy driven demixing picture breaks down. 

\section{Conclusions}
In conclusion, by performing computer simulations for binary mixtures of passive and eccentric active particles, we find that with increasing the eccentricity of the active particles, the gas-liquid like motility induced phase separation disappears, and the system undergoes a new dynamic demixing transition, in which active particles push passive particles forming a large cluster. We argue that the new dynamic demixing is due to the interplay between two different dynamics in the binary mixture. The circular-like trajectories of fast moving eccentric active particles, which can overlap with each other, deplete the slow moving passive particles, and when the density is high enough, to maximize the ``entropy'', the system demixes in analogy to the colloid-polymer mixtures~\cite{colloid_polymer,lekkerkerker1992,louis2000can,bolhuis2002i}. However, we find that the multi-body excluded volume effect between dynamic trajectories of active particles is very important, without which the demixing can not be reproduced in equilibrium using the obtained pair effective interaction. 
This makes the quantitative prediction on the dynamic demixing highly challenging and interesting for future study.
Moreover, we construct the phase diagram for the demixing of passive and eccentric active particles, and we find that the composition on the phase boundary of active rich phase changes non-monotonically with increasing the density of the system, which induces an intriguing re-entrant mixing with increasing the density of the system at certain fixed active-passive compositions.
Here we focused on the effect of eccentricity of active particles on the dynamic assembly of the system, and neglected hydrodynamic effects that
would be interesting for future experimental and theoretical studies~\cite{Yamamoto2013}. However, the results of this work can be also realized experimentally in the system of active granular matter, in which the hydrodynamic effect is neglectable, and the eccentricity of active particles can be better controlled~\cite{active_grain,active_grain2}. Therefore, our results open up a new way in designing active particles to collectively drive the assembly of passive matter~\cite{niprl2015,leonado2013,nisoftmatt2014,ni2013natcom}.

\begin{acknowledgments}
We thank Prof. Massimo Pica Ciamarra for helpful discussions. This work is supported by Nanyang Technological University Start-Up Grant (NTU-SUG: M4081781.120), the Academic Research Fund Tier 1 from Singapore Ministry of Education (M4011616.120, M4011873.120), and the Advanced Manufacturing and Engineering Young Individual Research Grant by the Science \& Engineering Research Council of Agency for Science, Technology and Research Singapore (M4070267.120).
We are grateful to the National Supercomputing Centre (NSCC) of Singapore for supporting the numerical calculations.
\end{acknowledgments}

\bibliographystyle{h-physrev}
\bibliography{reference}

\clearpage

\clearpage

\end{document}